\documentclass[aps,prl,twocolumn,amsmath,10pt,superscriptaddress,floatfix,nofootinbib,preprintnumbers]{revtex4-1}

\usepackage{mathrsfs}
\usepackage{amsfonts}
\usepackage{amsmath,amssymb,bm}
\usepackage{array}
\usepackage{verbatim}
\usepackage{epsfig}
\usepackage{graphicx}
\usepackage[colorlinks=true,citecolor=green,linkcolor=blue,urlcolor=blue]{hyperref}
\usepackage[normalem]{ulem}
\usepackage{xcolor}
\usepackage{ulem}
\usepackage{multirow}

\usepackage{graphicx}
\graphicspath{{./figs/}}
\usepackage{dcolumn}
\usepackage{bm}
\usepackage{slashed}
\usepackage{siunitx}
\usepackage{ulem,xpatch}
\usepackage{hyperref}
\usepackage[mathlines]{lineno}
\usepackage{comment}
\usepackage{soul}
\usepackage{xcolor}
\usepackage{xfrac}

\newcommand{\xb}{x_{\mbox{\tiny\!$B$}}}
\newcommand{\xhatb}{\hat{x}_{\mbox{\tiny\!$B$}}}
\newcommand{\xL}{x_{\mbox{\tiny $L$}}}
\newcommand{\zL}{z_{\mbox{\tiny $L$}}}
\newcommand{\Lqcd}{\Lambda_{\mbox{\tiny QCD}}}
\newcommand{\xh}{x_h}

\newcommand{\diff}[1]{\mathrm{d}#1}
\newcommand{\alfa}{\alpha}
\newcommand{\pTbar}{\overline{\rm p}_T}
\newcommand{\PTbar}{\overline{\rm P}_T}
\def\bea#1\eea{\begin{align}#1\end{align}}

\newcommand{\bef}{\begin{figure}[h!tb]\centering}
\newcommand{\beft}{\begin{figure}[t]\centering}
\newcommand{\eef}{\end{figure}}

\newcommand{\eref}[1]{(\ref{e.#1})}

\newcommand{\fref}[1]{Fig.~\ref{f.#1}}

\begin{document}

\title{Factorized approach to radiative corrections for inelastic lepton-hadron collisions}

\newcommand*{\SDU}{Key Laboratory of Particle Physics and Particle Irradiation (MOE), Institute of Frontier and Interdisciplinary Science, Shandong University, Qingdao, Shandong 266237, China}\affiliation{\SDU}
\newcommand*{\JLAB}{Theory Center, Jefferson Lab, Newport News, Virginia 23606, USA}\affiliation{\JLAB}
\newcommand*{\WM}{Department of Physics, The College of William \& Mary, Williamsburg, Virginia 23187, USA}\affiliation{\WM}

\author{Tianbo~Liu}\affiliation{\SDU}\affiliation{\JLAB}
\author{W.~Melnitchouk}\affiliation{\JLAB}
\author{Jian-Wei~Qiu}\affiliation{\JLAB}\affiliation{\WM}
\author{N.~Sato}\affiliation{\JLAB}

\preprint{JLAB-THY-20-3233}

\begin{abstract}
We propose a factorized approach to QED radiative corrections for inclusive and semi-inclusive deep-inelastic scattering to systematically account for QED and QCD radiation contributions to both processes on equal footing. This is achieved by utilizing factorization to resum logarithmically enhanced QED radiation into universal lepton distribution and fragmentation (or jet) functions. Numerical simulations suggest that the QED effects induced by the rotational distortion of the hadron transverse momentum, arising from the mismatch between the experimental Breit frame and the true photon-hadron frame, can be as large as 50\% for moderate $Q$, and become increasingly important for large transverse momenta. Our framework provides a uniform treatment of radiative effects for extracting three-dimensional hadron structure from high-energy lepton-hadron scattering at current and future facilities, such as the Electron-Ion Collider.
\end{abstract}
\maketitle

{\it Introduction ---}
Lepton-hadron deep-inelastic scattering (DIS) has played a critical role in the development of our understanding of the internal structure of nucleons and nuclei, since the first such experiments were performed at SLAC over 50 years ago~\cite{Bloom:1969kc}.
By measuring the momentum transfer, $q \equiv \ell-\ell'$, from an incident lepton $\ell$ scattered to a lepton with momentum $\ell'$, and keeping $Q \equiv \sqrt{-q^2} \gg 1/R$, where $R$ is the hadron radius, the DIS experiments provided a short-distance electromagnetic probe of the point-like quarks inside hadrons, ultimately giving birth to QCD as the theory of strong interactions.
Without observing specific final states other than the scattered lepton, this modern version of Rutherford scattering provided the first glimpse of the hadrons' internal landscape of quarks and gluons (or collectively, partons), parametrized through the parton distribution functions (PDFs) as probability densities for finding a parton inside the hadron with momentum fraction $x$~\cite{Feynman:1973xc}.

By detecting a hadron (or jet) of momentum $P_h$ in the final state, the semi-inclusive DIS (SIDIS) process has two naturally ordered momentum scales: $Q$, and the transverse momentum 
        $P_{hT} \equiv |\bm{P}_{hT}| \ll Q$,
defined in the ``photon-hadron'' frame, where the virtual photon collides with the hadron moving along the $z$-axis.
While the hard scale $Q$ localizes the probe to resolve the colliding parton and its momentum, the soft scale $P_{hT} \gtrsim 1/R$ provides the sensitivity needed to probe the parton's transverse motion inside the hadron.
With the leptonic plane defined by $\ell$ and $\ell'$ and the hadronic plane defined by $P$ and $P_h$, angular modulations between these planes in SIDIS allow the extraction of various transverse momentum dependent distributions (TMDs), which encode rich information about the hadron's three-dimensional structure in momentum space~\cite{Bacchetta:2006tn, Angeles-Martinez:2015sea, Collins:2016hqq, Diehl:2015uka, Gutierrez-Reyes:2019vbx, Liu:2018trl}.

\begin{figure}[t]
\includegraphics[width=0.99\columnwidth]{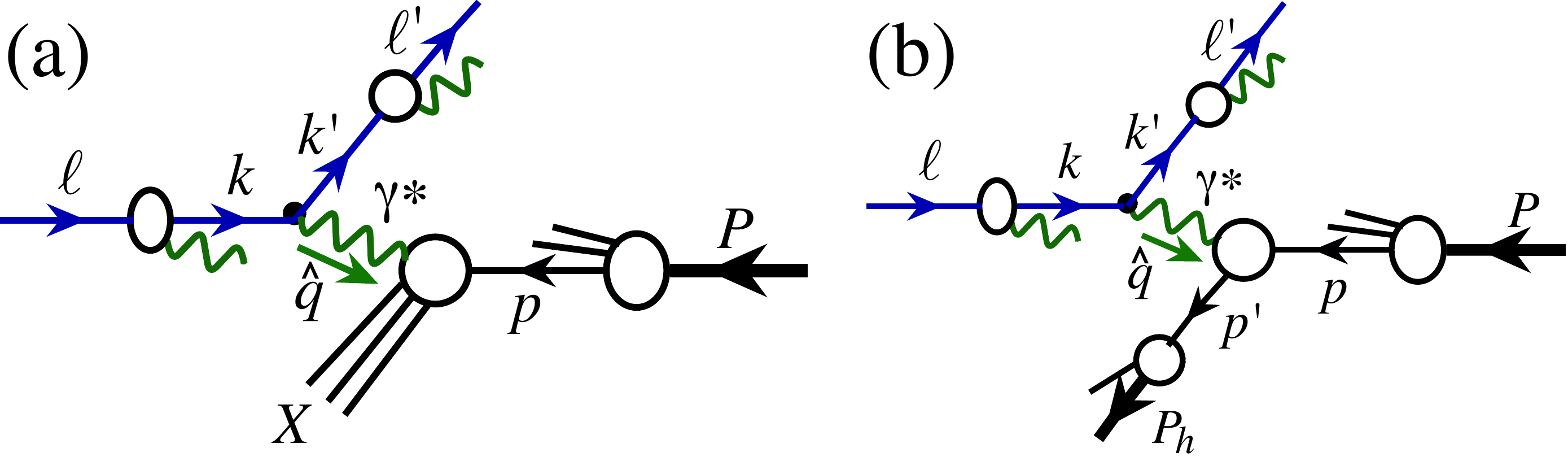} 
\caption{Scattering amplitudes for {\bf (a)} inclusive lepton-hadron DIS, and {\bf (b)} hadron (or jet) production in SIDIS.}
\label{f.fac}
\end{figure}

In practice, the collision with a large momentum transfer triggers radiation of photons from the colliding and scattered leptons and quarks [Fig.~\ref{f.fac}].
This radiation not only changes the momentum transfer~$q$, making it problematic to define $P_{hT}$ in the true photon-hadron frame and also alters the angular modulation between the leptonic and hadronic planes. 
Reliable extraction of PDFs and TMDs requires such collision-induced QED radiation to be taken into account. 
Historically, radiation induced modifications to inclusive DIS cross sections have been treated in the form of radiative corrections (RCs) to the Born cross section without radiation~\cite{Mo:1968cg, Bardin:1989vz, Badelek:1994uq}, and improved by resummation of logarithmic-enhanced RCs \cite{Kripfganz:1990vm, Spiesberger:1994dm, Blumlein:2002fy,Afanasev:2001zg}.  Similar RCs were also studied for $e^+ e^-$ collisions~\cite{Blumlein:2007kx, Frixione:2019lga, Bertone:2019hks, Ablinger:2020qvo}.
Without being able to account for all radiated photons experimentally, some of the RCs rely on measurement of the invariant mass of the hadronic final state and Monte Carlo simulation~\cite{Charchula:1994kf, Pierre:2019nry, Kwiatkowski:1990es, Arbuzov:1995id}.  
In practice, matching to the Born process by removing RCs becomes increasingly difficult beyond inclusive DIS~\cite{Ent:2001hm, Afanasev:2002ee, Akushevich:2019mbz}, and for more exclusive reactions with more particles observed it is effectively impractical.

In this paper, we propose a unified treatment of QED and QCD contributions to both inclusive DIS and SIDIS in a consistent factorization approach.
We present DIS as the inclusive production of a single lepton with large transverse momentum $\ell'_T \equiv |\bm{\ell}'_T| \gg \Lqcd$, and SIDIS as the production of a high-$\ell'_T$ lepton plus a large-$P_{hT}$ hadron (or jet) in lepton-hadron frame (see Fig.~\ref{f.fac}).
In the plane transverse to the lepton-hadron collision axis, the regime where $\bm{\ell}'_T$ and $\bm{P}_{hT}$ are almost back-to-back, namely,
    $\PTbar \equiv |\bm{\ell}'_T - \bm{P}_{hT}|/2 
    \gg |\bm{\ell}'_T + \bm{P}_{hT}| \equiv \pTbar$, 
is suited for TMD factorization, while the region where $\PTbar \sim \pTbar$ is suited for collinear (CO) factorization.

Our factorization approach to QED contributions to the DIS and SIDIS cross sections systematically organizes all order QED contributions into three separate categories:
{\bf (1)}~logarithmically enhanced collinear QED radiative contributions (which are resummed to all orders into universal lepton distribution functions (LDFs) and lepton fragmentation functions (LFFs)),
{\bf (2)}~infrared-safe contributions (as $m_e\to 0$) (which are calculated perturbatively as high-order corrections in powers of $\alpha$ to the short-distance hard parts), and 
{\bf (3)}~small corrections in inverse powers of the hard scale $Q$ (which are neglected).
Up to power corrections in ${\cal O}(m_e/Q)$, our QCD-like factorization approach provides a systematic and consistent way to include all order and complete QED radiative contributions to the observed cross sections, which is the main novelty and advantage of our factorization approach to QED contributions.
We will show that this approach especially impacts the extraction of TMDs from lepton-hadron scattering.

{\it Inclusive lepton-nucleon DIS ---} 
Starting with the more familiar case of inclusive DIS [Fig.~\ref{f.fac}(a)], 
in the QED Born approximation, and dropping mass terms, the spin-averaged cross section is given by
\begin{eqnarray}
\label{e.indis0} 
E' \frac{d\sigma_{\mbox{\tiny{\rm DIS}}}}{d^3 \ell'}
&\approx& \frac{4 \alfa^2}{s \xb y^2 Q^2}
\Big[ \xb y^2 F_1 + (1-y) F_2 \Big],
\end{eqnarray}
where $\alfa$ is the fine structure constant, 
$\xb = Q^2 / 2P\cdot q$ is the Bjorken variable, 
$s = (P + \ell)^2$, and 
\mbox{$y = P\cdot q / P\cdot \ell$}.
In this approximation, the structure functions $F_{1,2}(\xb,Q^2)$, which can be factorized in terms of PDFs with corrections suppressed by $1/Q^2$~\cite{Collins:1989gx}, would be cleanly extracted from DIS data~\cite{Chekanov:2001qu, Abramowicz:2015mha, Benvenuti:1989rh, Adams:1996gu, Arneodo:1996qe, Airapetian:2011nu, Tvaskis:2010as}.
In the presence of photon radiation, however, the exchanged photon momentum is
    $\hat{q} = k-k' \neq q$,
so that $F_{1,2}$ cannot be determined without accounting for all photon radiation.

Treating the QED and QCD contributions on the same footing, we can express the cross section in \eref{indis0} as~\cite{Nayak:2005rt}
\begin{eqnarray}
E'\frac{d\sigma_{\mbox{\tiny{\rm DIS}}}}{d^3 \ell'}
& \approx & 
\frac{1}{2s} \sum_{ija} 
\int_{\zL}^1 \frac{d\zeta}{\zeta^2} 
\int_{\xL}^1 \frac{d\xi}{\xi}\, D_{e/j}(\zeta)\, f_{i/e}(\xi)
\label{e.fac}\notag\\
& & \hspace*{0cm} \times
\int_{x_h}^1\frac{dx}{x} f_{a/N}(x)\,
\widehat{H}_{ia\to j}(\xi,\zeta,x),
\end{eqnarray}
where the indices $i,j,a$ include all QED and QCD particles~\cite{HQschemes},
and the dependence on the factorization scale $\mu$ is implicit.
The lower limits on the integrals in Eq.~\eref{fac} are given by
    \mbox{$\zL = \mbox{$(Q^2 - u)/s$}$}, 
    \mbox{$\xL = u/(Q^2 - \zeta s)$}
and \mbox{$\xh = \xi Q^2/(\xi\zeta s + u)$}, 
where \mbox{$u = (P - \ell')^2 = -(1 - y) s$}.
The LDF $f_{i/e}(\xi)$ gives the probability to find a lepton $i$ carrying a fraction $\xi$ of the incident lepton's momentum $\ell$~\cite{leptonstructure}, the LFF $D_{e/j}(\zeta)$ describes the emergence of the final lepton with momentum $\ell'$ from lepton $j$ of momentum~$\ell'/\zeta$, and $f_{a/N}(x)$ is the nucleon PDF with momentum fraction $x$ carried by the colliding parton $a$.
Both the LDF and LFF are defined in analogy with the quark PDF in the nucleon and the quark to hadron fragmentation function (FF)~\cite{Collins:1981uw}, respectively.

Unlike the PDFs in QCD, the LDFs and LFFs are calculable perturbatively in QED. 
Focusing for brevity on the ``valence'' lepton contribution [$i=j=e$]
and leading logarithmic contribution, at leading order (LO) in $\alpha$ we have $f^{(0)}_{e/e}(\xi) = \delta(\xi-1)$, while at next-to-leading order (NLO) in the $\overline{\rm MS}$ scheme,
\begin{eqnarray}
f_{e/e}^{(1)}(\xi,\mu^2)
= \frac{\alpha}{2\pi}
\left[
    \frac{1+\xi^2}{1-\xi}
    \ln\frac{\mu^2}{(1-\xi)^2 m^2_e}
\right]_+ , 
\label{e.lpdf1}
\end{eqnarray}
where $m_e$ is the lepton mass and the standard ``+'' prescription is used. 
Similarly, for the LFFs, one has $D^{(0)}_{e/e}(\zeta) = \delta(\zeta - 1)$ at LO, while at ${\cal O}(\alpha)$ in the $\overline{\rm MS}$ scheme,
\begin{eqnarray}
D_{e/e}^{(1)}(\zeta,\mu^2)
= \frac{\alpha}{2\pi}
\left[
    \frac{1+\zeta^2}{1-\zeta} 
    \ln \frac{\zeta^2\mu^2}{(1-\zeta)^2 m_e^2}
\right]_+ .
\label{e.lff1}
\end{eqnarray}
As with PDFs and FFs, the logarithmic-enhanced high-order contributions to LDFs and LFFs can be systematically resummed by solving the corresponding QED evolution equations~\cite{Williams:1934ad, vonWeizsacker:1934nji, Dokshitzer:1977sg, Gribov:1972ri, Lipatov:1974qm, Altarelli:1977zs}. 
 
In Eq.~\eref{fac}, $\widehat{H}_{ia \to j}$ is the lepton-parton scattering cross section, with all logarithmic CO sensitivities along the direction of observed momenta, $\ell$, $\ell'$ and $P$, removed.
It can be calculated perturbatively to ${\cal O}(\alpha^m \alpha_s^n)$ by applying \eref{fac} to a point-like parton ($q$ or $g$) state, and at LO in both QED and QCD can be written as
\begin{eqnarray}
\widehat{H}_{eq \to e}^{(2,0)}
&=& \frac{4\alpha^2 e_q^2}{Q^4}\, 
\frac{x^2 \zeta\, \big[ (\xi\zeta s)^2 + u^2\big]}
     {\xi^2\, (\xi\zeta s + u)}\,
\delta\big( x - \xh \big).
\label{e.H20}
\end{eqnarray}
Substituting \eref{H20} into Eq.~\eref{fac} and choosing $f_{e/e} \approx f_{e/e}^{(0)}$ and $D_{e/e} \approx D_{e/e}^{(0)}$, one can reproduce the lepton-nucleon cross section in \eref{indis0} by noting that at ${\cal O}(\alpha_s^0)$ the structure functions 
    $F_2(\xb) = 2 \xb F_1(\xb) = \sum_q e_q^2\, \xb f_{q/N}(\xb)$.

The advantage of the factorized cross section in \eref{fac} is that it allows resummation of large QED contributions collinearly sensitive to the incident lepton into $f_{i/e}$, along with those collinearly sensitive to the scattered lepton into $D_{e/j}$ and those collinearly sensitive to the colliding nucleon into $f_{a/N}$, by adding QED corrections to their evolution kernels \cite{Martin:2004dh,Ball:2013hta}
and infrared safe contributions to $\widehat{H}_{ia \to j}$ order-by-order in powers of $\alpha$.
With photon radiation, the invariant mass of the exchanged virtual photon in Fig.~\ref{f.fac}(a) is 
    $\widehat{Q}^2 \equiv - \hat{q}^2 = (\xi/\zeta)\, Q^2$, 
with a minimum value 
    $\widehat{Q}^2_{\rm min} = (1-y)/(1-y\, \xb)\, Q^2 \ll Q^2$
if $y$ is large and $\xb$ is small. 
The factorization in \eref{fac} requires that $\widehat{Q}^2_{\rm min}\gg \Lqcd^2$, which could reduce the kinematic reach to the regime of small $\xb$ and large $y$ planned at the future Electron-Ion Collider (EIC)~\cite{Accardi:2012qut}.

The numerical impact of RCs on inclusive DIS is illustrated in \fref{sigmaratio} for the ratio of
    $\sigma_{{\rm no\,}\mbox{\tiny {\rm RC}}} 
    \equiv E' d\sigma_{\mbox{\tiny{\rm DIS}}}/d^3 \ell'$,
evaluated with LO LDFs and LFFs, 
to the cross section with radiation, $\sigma_{\mbox{\tiny {\rm RC}}}$,
evaluated with NLO LDFs and LFFs (``NLO'' in \fref{sigmaratio}) and with 
these evolved (``RES'' in \fref{sigmaratio}), for $\mu_0=m_e$.
We note that the choice of $\mu_0$ is not unique, which impacts the size of uncalculated higher-order contributions to LDFs and LFFs, and will be explored further in future work.

For the PDFs in the nucleon we use the recent JAM parametrization~\cite{Sato:2019yez}, although the ratios with other PDF sets~\cite{Accardi:2016qay} are almost indistinguishable.
We choose the scale $\mu^2 = \max[\, m_c^2, \ell'^2_T\, ]$, with $m_c=1.28$~GeV, 
but find weak dependence on the scale.

\begin{figure}[t]
    \includegraphics[width=0.99\columnwidth]{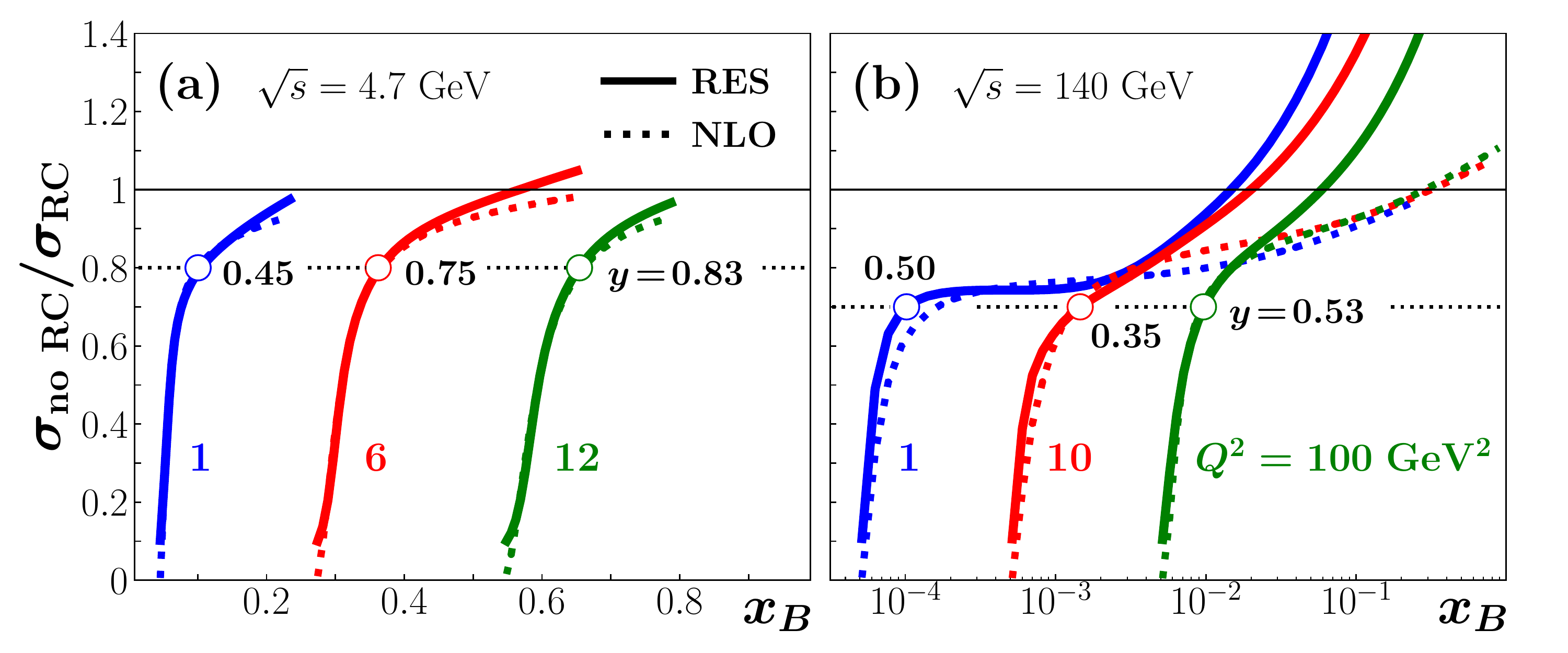}
    \vspace*{-0.6cm}
    \caption{Ratio of inclusive $ep$ cross sections with no RCs to those with RCs versus $\xb$ at fixed $Q^2$ for
    {\bf (a)} Jefferson Lab ($\sqrt{s}=4.7$~GeV) and 
    {\bf (b)} EIC ($\sqrt{s}=140$~GeV) kinematics, using the RES and NLO schemes.
    The values of $y$ at which the RCs exceed 20\% (30\%) for the Jefferson Lab (EIC) kinematics (dotted lines) are indicated by the open circles.}
\label{f.sigmaratio}
\end{figure}

The ratios for cross sections at kinematics typical of Jefferson Lab experiments and those planned for the EIC  in \fref{sigmaratio} show dramatic effects of RCs in certain regions of phase space. 
The general trend is 
    $\sigma_{\mbox{\tiny {\rm RC}}} > \sigma_{{\rm no\,}\mbox{\tiny {\rm RC}}}$ 
at lower $\xb$, corresponding to smaller $\ell'_T$, which is the effective hard scale for $\gamma^* N$ scattering, giving more phase space for photon radiation.
At larger $\xb$, where $\ell'_T$ increases to its kinematic limit, radiation requires the active quark to have a larger momentum fraction, and we have
    $\sigma_{\mbox{\tiny {\rm RC}}} < \sigma_{{\rm no\,}\mbox{\tiny {\rm RC}}}$. 
Note that a cut on the final state hadronic mass ($> 2$~GeV) is made to avoid the nucleon resonance region, which restricts the maximum $\xb$ that can be reached at fixed $Q^2$.
The results in \fref{sigmaratio} clearly indicate that extreme care should be taken when extracting partonic information from inclusive DIS cross sections at low $\xb$ and high $Q^2$ values.
For less inclusive observables, the effects can be even more dramatic, as we discuss next.

{\it Semi-inclusive DIS ---}
With photon radiation, not only can the exchanged photon's invariant mass squared, $\widehat{Q}^2$, be very different from $Q^2$, but its direction $\bm{\hat{q}}$ in \fref{fac}(b) can also be modified from that of $\bm{q}$.
Without having all induced photon radiation accounted for, one {\it cannot} fully determine the photon-nucleon frame, which prevents identifying precisely the $P_{hT} \ll \widehat{Q}$ region needed for TMD factorization~\cite{Ji:2004wu, Aybat:2011zv, Aybat:2011ge, Bacchetta:2004jz}.

In analogy with the treatment of DIS as the inclusive production of a single lepton at large $\ell'_T$, we define SIDIS as the inclusive production of a high-$\ell'_T$ lepton plus a high-$P_{hT}$ hadron (or jet) in the lepton-hadron frame,
    $d\sigma_{\mbox{\tiny{\rm SIDIS}}}/dy_{\ell'} d^2\bm{\ell}'_T\, dy_h d^2\bm{P}_{hT}$, 
where both $\ell'_T$ and $P_{hT}$, and their rapidities $y_{\ell'}$ and $y_h$, are well-defined and measured.  
In the plane transverse to the lepton-nucleon collision, the scattered lepton and produced hadron are generally back-to-back, $\PTbar \gg \pTbar$, and the SIDIS cross section is suited for TMD factorization.
The $\PTbar$ defines the scale of the hard collision, while the momentum imbalance, $\pTbar$, is generated by the induced QED and QCD radiation on top of active particles' intrinsic transverse momenta.

Like QCD radiation that generates transverse momentum broadening of the active parton, QED radiation can lead to a transverse momentum broadening of the active election, which contributes to the observed momentum imbalance $\overline{\rm p}_T$.  
After a careful study of the QED transverse momentum broadening in terms of transverse momentum dependent LDFs and Sudakov resummation of QED logarithms, we found that the broadening from QED radiation is much smaller than the corresponding broadening from QCD radiation, mainly due to the fact that $\alpha \ll \alpha_s$~\cite{LMQS}. 
We conclude, therefore, that the effects of QED radiation on SIDIS can be safely treated in CO factorization. 
The resulting combined QED and QCD factorization for SIDIS then reads 
\begin{align}
\frac{d\sigma_{\mbox{\tiny{\rm SIDIS}}}^h}
     {dy_{\ell'} d^2\bm{\ell}'_T\, dy_h d^2\bm{P}_{hT}}
&\approx \sum_{ij}
\int\frac{d\zeta}{\zeta^2} \int d\xi\, D_{e/j}(\zeta)\, f_{i/e}(\xi)
\nonumber\\
&\times
\frac{d\hat{\sigma}_{\mbox{\tiny{\rm SIDIS}}}^{ij, h}}
     {dy_{k'} d^2\bm{k}'_T\, dy_h d^2\bm{P}_{hT}}\, ,
\label{e.sidis}
\end{align}
where $d\hat{\sigma}_{\mbox{\tiny{\rm SIDIS}}}^{ij, h}$ is the partonic SIDIS cross section with all perturbative CO sensitivities along the direction of lepton momentum $\ell$ and $\ell'$ removed.

When $\PTbar \gg \pTbar$, the partonic cross section $d\hat{\sigma}_{\mbox{\tiny{\rm SIDIS}}}^{ij, h}$ can be evaluated in terms of TMD factorization, which is defined and proved in the photon-nucleon frame. 
If $\widehat{Q}^2 \gg \Lqcd^2$, we can approximate 
    $d\hat{\sigma}_{\mbox{\tiny{\rm SIDIS}}}^{ij, h}$
at lowest order in QED by one-photon exchange.
Then, for any given pair of $(\xi,\zeta)$, we have a well-defined virtual photon-nucleon frame given by $\hat{q}$, and can compute the Lorentz invariant cross section 
    $d\hat{\sigma}_{\mbox{\tiny{\rm SIDIS}}}^{ij, h}$
in Eq.~(\ref{e.sidis}) in two steps: \\
\hspace*{0.2cm} {\bf (i)}
evaluate $d\hat{\sigma}_{\mbox{\tiny{\rm SIDIS}}}^{ij, h}$ in the virtual photon-nucleon frame (for a given $(\xi,\zeta)$) in terms of TMD factorization and the corresponding momentum variables~\cite{Bacchetta:2006tn}; \\
\hspace*{0.2cm} {\bf (ii)}
apply a $(\xi,\zeta)$-dependent Lorentz transformation to change all momentum variables of $d\hat{\sigma}_{\mbox{\tiny{\rm SIDIS}}}^{ij, h}$ in the virtual photon-nucleon frame to a frame suitable for comparison with experiment (such as the lepton-nucleon frame or experimentally defined Breit frame).

When $\PTbar \sim \pTbar$, $d\hat{\sigma}_{\mbox{\tiny{\rm SIDIS}}}^{ij, h}$ should be evaluated in terms of CO factorization in the lepton-hadron frame. 
The matching between the regimes where these two factorizations are applicable has been a topic of considerable interest~\cite{Collins:2016hqq, Collins:1984kg, Collins:2017oxh, Gamberg:2017jha, LMQS}. 
With the CO approximation for the QED contribution and one-photon exchange, we could perform both the calculation and the matching of the two regimes in the virtual photon-nucleon frame for a given $(\xi,\zeta)$, and then apply a Lorentz transformation to change the variables in the virtual photon-nucleon frame to a physical frame where the comparison with experimental data is performed.

To demonstrate the impact of the QED radiation explicitly, we consider the case of the unpolarized SIDIS cross section with $P_{hT}$ defined in the Breit frame --- the photon-nucleon frame without photon radiation~\cite{Bacchetta:2006tn},  
\begin{eqnarray}
\label{e.sidis0}
&&\frac{d\sigma_{\mbox{\tiny{\rm SIDIS}}}^h}{d\xb dy\, dz\, dP^2_{hT}}
= \int_{\zeta_{\rm min}}^1 \! d\zeta
  \int_{\xi_{\rm min(\zeta)}}^1 \!\! d\xi\,
  D_{e/e}(\zeta)\, f_{e/e}(\xi) 
\notag\\   
&&{\hskip -0.1cm}
\times 
  \left[ \frac{\xhatb}{\xb\, \xi \zeta} \right]
  \left[
    \frac{(2\pi)^2\, \alpha}{\xhatb \hat{y}\, \widehat{Q}^2}
    \frac{\hat{y}^2}{2(1-\hat{\varepsilon})}
    F_{UU}^h(\xhatb,\widehat{Q}^2,\hat{z},\widehat{P}_{hT})
  \right]
\end{eqnarray}
where $z = P\cdot P_h / P\cdot q$ and $\varepsilon = (1-y) / (1-y+y^2/2)$.
All variables with a ``{\small $\wedge$}'' are defined in the virtual photon-nucleon frame, which can be converted to the Breit frame by a rotation in the nucleon's rest frame, or in general by a Lorentz transformation.
In Eq.~(\ref{e.sidis0}) the expression in the first bracket [\,$\cdots$] is the Jacobian for transforming variables between the two frames, which $\to 1$ as $\xi$, $\zeta \to 1$; the second bracket [\,$\cdots$] is the result without QED radiation.
At LO in $\alpha_s$, the SIDIS structure function is given by~\cite{Bacchetta:2006tn}
\begin{eqnarray}
F_{UU}^h
&=& \xb \sum_q e_q^2 \int d^2\bm{p}_T\, d^2\bm{k}_T\,
    \delta^{(2)}\big( \bm{p}_T - \bm{k}_T - \bm{q}_T \big) 
\nonumber\\
& & \hspace*{1.5cm} \times\ 
    f_{q/N}(\xb,\bm{p}_T^2)\, D_{h/q}(z,\bm{k}_T^2),
\label{e.tmds}
\end{eqnarray}
where $\bm{q}_T = \bm{P}_{hT}/z$.
Using a Gaussian {\it ansatz} for the transverse momentum dependence of the TMDs, with transverse widths
    $\langle \bm{p}_T^2 \rangle = 0.57(8)~{\rm GeV}^2$ and  
    $\langle \bm{k}_T^2 \rangle = 0.12(1)~{\rm GeV}^2$~\cite{Anselmino:2013lza}, 
we show in \fref{sidis_bf}(a) the impact of photon radiation for SIDIS at EIC energies with $\sigma_{\mbox{\tiny {\rm RC}}}$ given by (\ref{e.sidis0}), and $\sigma_{{\rm no\,}\mbox{\tiny {\rm RC}}}$ given by the same expression but evaluated with 
LO LDF and LFF.
While the validity of the TMD formalism is limited to the range $q_T/Q < 1$, we extend the calculation up to $q_T/Q = 2$ to use as a proxy for illustrating the QED effects also at large $q_T$.
A more precise formulation that includes proper TMD evolution matched to CO factorization at large $P_{hT}$ is left for future work.

The resulting cross section ratios in \fref{sidis_bf}(a) indicate that the QED effects can be as large as 50\% for moderate $Q$, and become increasingly important for larger transverse momenta.
In addition, by removing the QED rotational effects from the transverse momentum $\widehat{P}_{hT}$ one sees that most of the QED effect is encoded in the rotational distortion of the transverse momentum, as quantified by the difference between the solid and dashed lines.
This demonstrates clearly that the experimental Breit frame and the true photon-nucleon frame needed for QCD factorization do not in general coincide, indicating the need to treat QED and QCD on the same footing in studies of hadron structure in SIDIS.

\begin{figure}[t]
    \hspace*{-0.2cm}\includegraphics[width=0.245\textwidth]{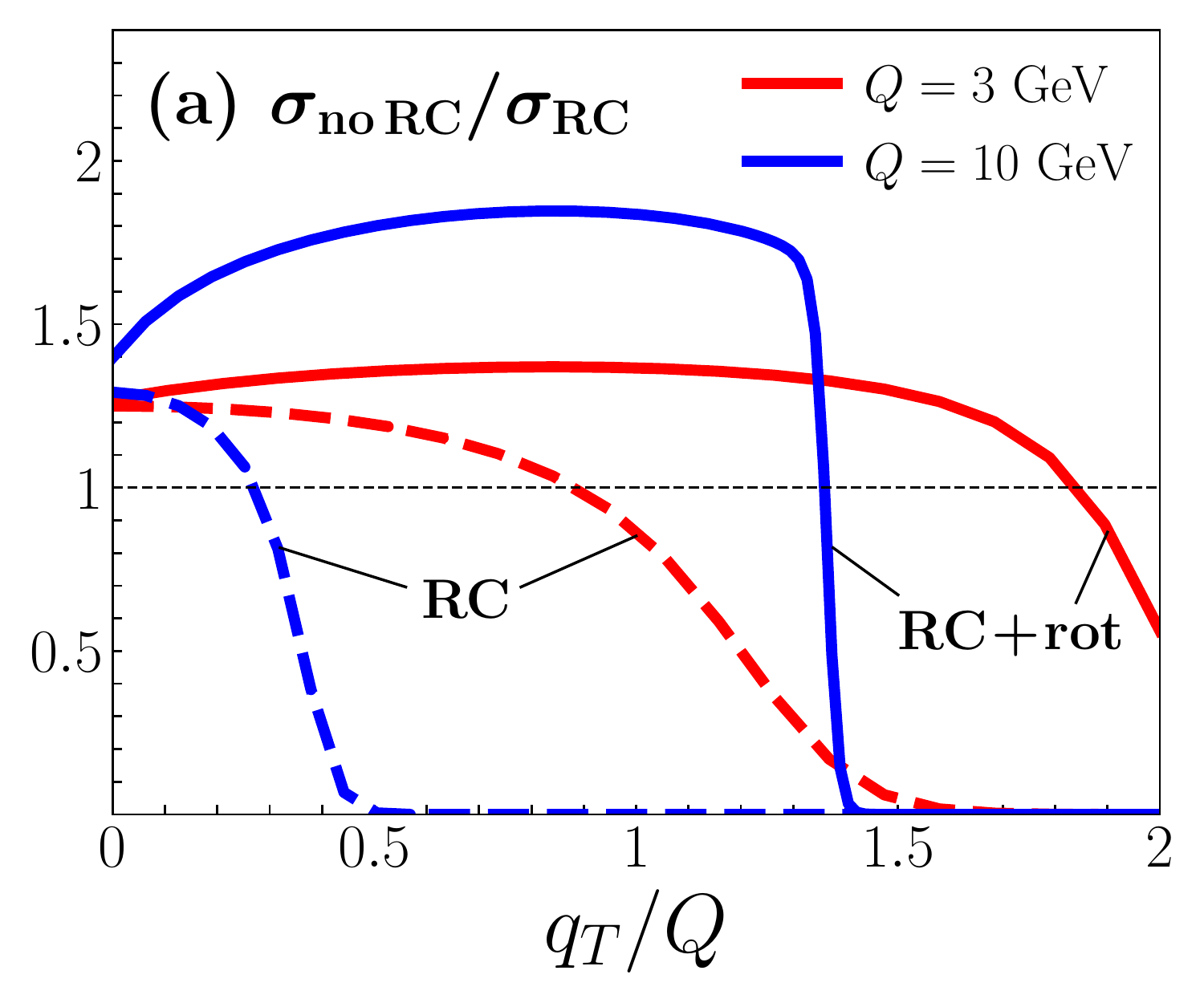}
    \hspace*{-0.2cm}\includegraphics[width=0.245\textwidth]{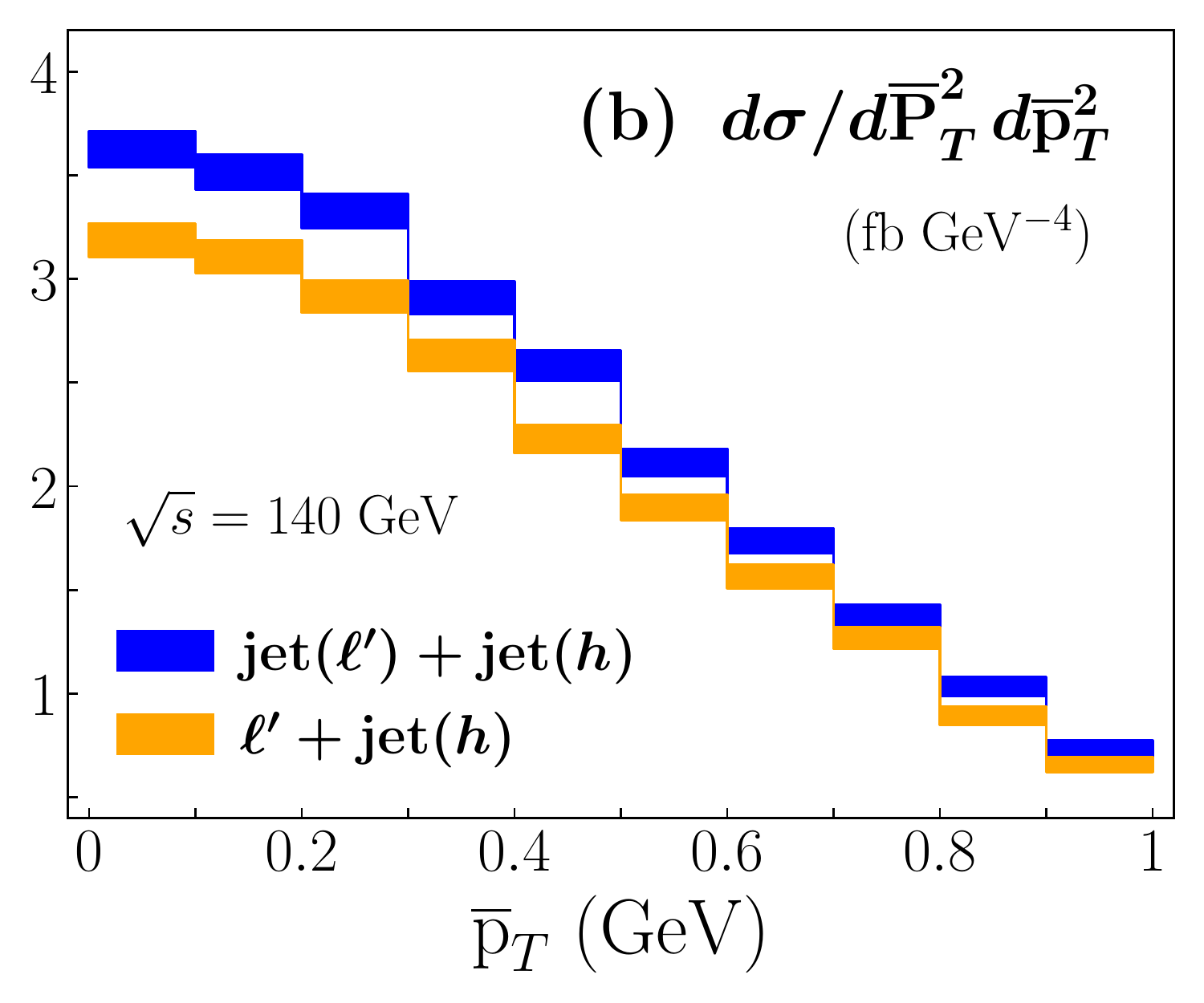}
    \vspace*{-0.1cm}
    \caption{QED effects on SIDIS cross sections at $\sqrt{s}=140$~GeV:
    {\bf (a)} Ratio without to with RCs in the photon-nucleon frame versus $q_T/Q$ for $y=0.4$ and $z=0.5$, with QED RCs including rotation distortion of hadron transverse momentum (solid) and without rotation (dashed).
    {\bf (b)} Lab frame SIDIS cross section with QED effects versus $\pTbar$ for $\PTbar \in [9,10]$~GeV and $|y_{\ell'}, y_h| \leq 3.5$, with lepton jet + hadron jet (blue) and single lepton + hadron jet (orange) final states.}
\label{f.sidis_bf}
\end{figure}

With the collinear factorization for the QED radiation in (\ref{e.sidis}) and the one-photon approximation, we can avoid going through the virtual photon-nucleon frame by formulating  SIDIS directly in the lepton-hadron frame, if we observe a final-state {\it jet} instead of a hadron.  
At LO, taking $D_{\tiny{\rm jet}/q}(z) \approx \delta(z-1)$, we then have
\begin{eqnarray}
\hspace*{-0.7cm}
\frac{d\sigma_{\mbox{\tiny{\rm SIDIS}}}^{\rm jet}}
     {dy_{\ell'}\, dy_h\, d^2{\overline{\bf P}_T}\, \diff^2{\overline{\bf p}_T}}
&=& \sum_q e_q^2
  \frac{2 \alpha^2}{\hat{s}^2}
  \int_{\zeta_{\rm min}}^1 
  \frac{d\zeta}{\zeta^2}\, D_{e/e}(\zeta)
\nonumber\\
& & \hspace*{-1.0cm} \times\
  \big[\xi f_{e/e}(\xi)\big]
  \big[x \widetilde{f}_{q/N}(x,\bm{k}_T^2)\big]\,
  \frac{\hat{s}^2 + \hat{u}^2}{\hat{t}^2},
\label{e.sidispT}
\end{eqnarray}
where 
$\{ \xi, x \} = \big[ (\ell'_T/\zeta)\, e^{\pm y_{\ell'}}
                    + P_{hT}\, e^{\pm y_h}
                \big]/\sqrt{s}$,
$\hat{s} = x \xi s$, 
$\hat{t} = -(\xi \sqrt{s}/\zeta)\, \ell'_T e^{-y_{\ell'}}$,
and
$\hat{u} = -\xi \sqrt{s}\, P_{hT}\, e^{-y_h}$.
From the kinematic constraints $x \leq 1$ and $\xi \leq 1$, one has
    $\zeta_{\rm min}
    = {\rm max}\big\{ \ell'_T\, e^{\pm y_{\ell'}}/(\sqrt{s} - P_{hT} e^{\pm y_h}) \big\}$.
In Eq.~(\ref{e.sidispT}), the partonic transverse momentum of the active quark, 
    $\bm{k}_T = \bm{\ell}'_T/\zeta + \bm{P}_{hT}$, 
is directly responsible for the momentum imbalance between the scattered lepton and the observed jet.
Moreover, if one detects a {\it lepton jet} in the final state, corresponding to $\zeta \to 1$, $\bm{k}_T$ would be determined entirely by $\bm{\ell}'_T$ and $\bm{P}_{hT}$, as $\bm{k}_T = \overline{\bf p}_T$, and the cross section \eref{sidispT} would become directly proportional to the TMD PDF $\widetilde{f}_{q/N}$.
In this case the hard scale is given by
    \mbox{$\PTbar \equiv |\bm{\ell}'_T - \bm{P}_{hT}|/2$},
as defined above.

By integrating over rapidities $|y_{\ell', h}| \leq 3.5$, and for a typical bin $\PTbar \in [9,10]$~GeV, in Fig.~\ref{f.sidis_bf}(b) we plot the SIDIS cross section in \eref{sidispT} as a function of the momentum imbalance $\pTbar$ at EIC kinematics.
In the numerical calculation we take the same factorized {\it ansatz} for the TMD PDF.  
As expected, the momentum imbalance directly follows the transverse momentum profile of the TMD PDF.  
If instead of a lepton jet we were to observe an individual electron, the electron FF $D_{e/e}$ in \eref{sidispT} would produce a distorted spectrum slightly lower in magnitude and enhanced at large $\overline{\rm p}_T$ due to the presence of the LDF, as illustrated in \fref{sidis_bf}(b).
Note that at LO, the differential SIDIS cross section \eref{sidispT} for a hadronic jet in the final state would {\it not} exist without QED radiation, which renders the LDF $f_{e/e}(\xi)$ finite.

{\it Outlook ---}
The unified factorized approach to QED and QCD contributions to inclusive DIS and SIDIS presented here has important implications for future analyses of hard scattering at the EIC.
We stress that even though $\alpha$ is very small, the logarithmic enhanced QED RCs could significantly alter the momentum transfer to the colliding nucleon.
Without being able to account for all radiation, the photon-hadron frame is not well-defined and the standard $\xb$ and $Q^2$ variables do not fully control the momentum transfer, which could impact the precision with which TMDs can be extracted.

Our approach allows the systematic resummation of the logarithmically enhanced RCs into factorized LDFs and LFFs that are universal for all final states, applicable for DIS, SIDIS, as well as for $e^+ e^-$ annihilation and Drell-Yan lepton-pair production processes, leaving the fixed-order QED corrections completely IR-safe and stable in the limit as $m_e \to 0$. 
It provides an alternative paradigm for a uniform treatment of QED RCs for the extraction of PDFs, TMDs and other partonic correlation functions in the quest to map the nucleon's three-dimensional structure in momentum space from lepton-hadron collision data.\\

{\it Acknowledgments ---}
We thank the participants of the {\it Theory-Experiment Dialogue} at Jefferson Lab, including A.~Accardi, H.~Avakian, \mbox{J.-P.} Chen, R.~Ent, C.~E.~Keppel, A.~Prokudin, T.~C.~Rogers and P.~Rossi, for helpful discussions.
This work is supported by the U.S. Department of Energy contract DE-AC05-06OR23177, under which Jefferson Science Associates, LLC, manages and operates Jefferson Lab, and within the framework of the TMD Topical Collaboration. 
The work of TL is supported in part by National Natural Science Foundation of China under Contract No. 12175117.
The work of NS was supported by the DOE, Office of Science, Office of Nuclear Physics in the Early Career Program.


\end{document}